\documentclass[twoside]{mhd}
\usepackage{graphicx}
\usepackage{bm}
\mhdhead{40}{1}{1}	
\title{Liquid metal experiments on the\\ helical magnetorotational instability}
\author{ F. Stefani\inst{1}, G. Gerbeth\inst{1}, Th. Gundrum\inst{1},\\ J. Szklarski\inst{2},
G. R\"udiger\inst{3}, R. Hollerbach\inst{4}}
\institute{Forschungszentrum Dresden-Rossendorf, P.O. Box 510119,
           D-01314 Dresden, Germany
     \and Institute of Fundamental Technological Research, 
      21, \'{S}wi\c{e}tokrzyska , 00-049 Warsaw, Poland
       \and Astrophysikalisches Institut Potsdam, An der Sternwarte 16, D-14482 Potsdam, Germany
       \and Department of Applied Mathematics, University of Leeds, Leeds, LS2 9JT, UK
}
\begin{document}
\maketitle
\begin{abstract}
The magnetorotational instability (MRI) plays an essential role in 
the formation of stars and black holes. By destabilizing hydrodynamically 
stable Keplerian flows, the MRI triggers turbulence and enables outward 
transport of angular momentum in accretion discs. We present the results 
of a liquid metal Taylor-Couette experiment under the influence of helical 
magnetic fields that show typical features of MRI at Reynolds numbers of 
the order 1000 and Hartmann numbers of the order 10. Particular focus 
is laid on an improved experiment in which split end caps are 
used to minimize the Ekman pumping.
\end{abstract}

\section*{Introduction.}      

Magnetic fields play a double role in the cosmos:  First, planetary, 
stellar, and galactic fields are produced by the homogeneous dynamo effect 
in moving electrically conducting fluids. Second, magnetic fields can 
accelerate tremendously the formation of stars and black holes by 
enabling outward directed transport of angular momentum in accretion disks by 
virtue of the so-called magnetorotational instability (MRI). 
This instability was discovered as early as 1959, when Velikhov 
showed that a Taylor-Couette flow in its hydrodynamically stable regime 
(i.e. with outward increasing angular momentum) can be destabilized by an 
applied axial magnetic field \cite{VELIKHOV}. It was only in 1991, however, 
that the importance of MRI for accretion disks in the vicinity of young stars 
and black holes was realized in a seminal paper by Balbus and Hawley 
\cite{BAHA}. 

The last decades have seen tremendous theoretical and computational progress 
in understanding the dynamo effect and the MRI. The 
hydromagnetic dynamo effect has even been verified experimentally in 
large-scale liquid sodium 
facilities in Riga, Karlsruhe, and Cadarache, and is presently studied in 
laboratories around the world \cite{RMP,ZAMM}. In contrast, 
attempts to study the MRI in the laboratory have been less successful so far. 
Recently, an MRI-like instability has been observed on the background of a 
turbulent 
spherical Couette flow \cite{SISAN}, but the genuine idea that MRI 
would destabilize a hydrodynamically stable flow was not realized in this
experiment.

One of the basic problems for the experimental investigation of the 
''standard MRI'', with  only an axial magnetic field being externally 
applied, is the need for flows with large magnetic Reynolds numbers 
$Rm$. The crucial point is that the azimuthal field, which is essential
for  the MRI mechanism to work, must be produced from the 
applied axial field by induction effects proportional to $Rm$. The 
natural question, why not substitute the induction process by externally 
applying the azimuthal field as well, was addressed by Hollerbach 
and R\"udiger \cite{HORU}. 
Exemplified in a Taylor-Couette configuration, it was shown 
that the scaling properties for 
this ''helical MRI'', as we call it now, are completely 
different from those of the ''standard MRI''. While the latter is 
governed by the magnetic Reynolds number $Rm$ and the Lundquist 
number $S$, the 
former depends only on the hydrodynamic Reynolds number $Re$ and on the 
Hartmann number $Ha$. For liquid metals, with their small 
magnetic Prandtl number $Pm$, this makes a dramatic difference 
for the feasibility of MRI experiments, since $Rm=Pm Re$ and 
$S=Pm^{1/2}Ha$.

\begin{figure}[t]
\begin{center}
\unitlength=\textwidth
\includegraphics[width=0.95\textwidth]{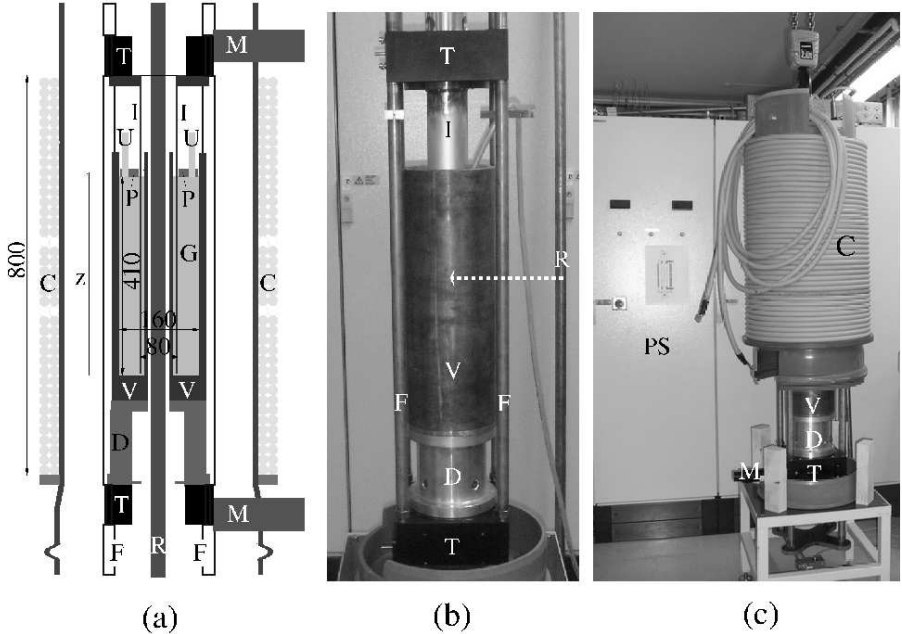}
\caption{The PROMISE experiment. (a) Sketch. 
(b) Photograph of the central part. 
(c) Total view with the coil being installed. 
V - Copper vessel, I - Inner cylinder, 
G - GaInSn, U - Two ultrasonic transducers, 
P  - Plexiglas lid, T - High precision turn-tables, 
M - Motors, F - Frame, C - Coil, 
R - Copper rod, 
PS - Power supply for currents up to 8000 A.  
The indicated dimensions are in mm.}
\end{center}
\end{figure}

We have to point out, though, that there is no real 
breach between ``standard MRI" and ``helical MRI". This can be
clearly seen in Fig. 1 of \cite{HORU},  where the critical
$Rm$ is plotted against the rotation ratio $\mu:=f_{\rm o}/f_{\rm i}$ 
of outer to inner cylinder of a Taylor-Couette set-up.
The extremely steep increase of this curve
at the Rayleigh line, which occurs for a purely axial magnetic field,  
is just smeared out when an azimuthal field is added. 

In this paper, we summarize the main results of a 
first experimental 
verification 
of this idea obtained in the framework of the PROMISE experiment 
(Potsdam ROssendorf Magnetic InStability Experiment) which were 
already published in \cite{PRL,APJL,NJP,PROMISEAN}, and we report some 
new results of an improved version of this experiment.

\section{The experimental facility.}

The basic part of PROMISE is a cylindrical containment vessel 
V made of copper (see Fig. 1). The inner wall of the vessel V is 
10 mm thick, and extends in radius from 22 to 32 mm; the outer 
wall is 15 mm thick, extending from 80 to 95 mm.  This vessel is 
filled with the eutectic alloy Ga$^{67}$In$^{20.5}$Sn$^{12.5}$ which 
is liquid at room temperatures.
The vessel V, which is completely made of copper in the first version 
of the experiment (henceforth called PROMISE 1), 
is fixed, via an aluminum spacer D, on a precision turntable T; 
the outer copper cylinder of the vessel represents the outer 
cylinder of the Taylor-Couette cell. The inner copper cylinder 
I of the Taylor-Couette flow is fixed to an upper turntable, and 
is immersed into the liquid metal from above. It has a thickness of 4 mm, 
extending in radius from 36 to 40 mm, thus leaving a gap 
of 4 mm between this immersed cylinder I and the inner wall 
of the containment vessel V.  The actual Taylor-Couette 
cell extends in radial direction over a cylindrical gap 
of width $d=r_{\rm o}-r_{\rm i}=40$ mm, and in axial direction over the liquid metal 
height $h=400$ mm, resulting in an aspect ratio of 10.

\begin{figure}[t]
\begin{center}
\unitlength=\textwidth
\includegraphics[width=0.95\textwidth]{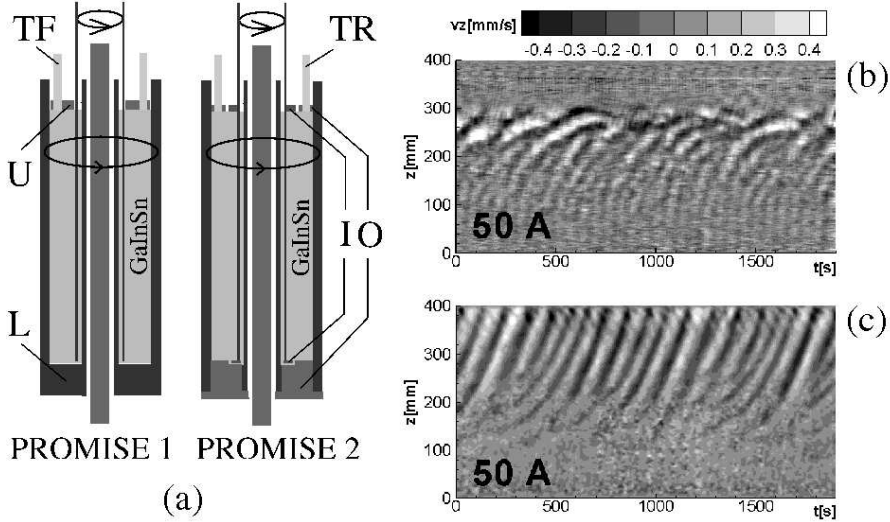}
\caption{(a) The asymmetric end cap configuration in 
PROMISE 1 is replaced by a symmetric one in PROMISE 2. 
U - Upper end cap fixed to the frame, L - Lower copper end cap rotating 
with outer cylinder, 
I -  Inner rings, R - Outer rings, 
TF - Fixed ultrasonic transducer, TR - Rotating ultrasonic transducer. 
(b) and (c) Axial velocity perturbation 
for $f_{\rm i}=0.06$ Hz (i.e. $Re=1779$), 
$f_{\rm o}=0.0162$ (i.e. $\mu=0.27$), $I_{\rm coil}=50$ A (i.e. Ha=7.9).
(b) PROMISE 1  with $I_{\rm rod}=6000$ A (i.e. $\beta=5.9$). (c) PROMISE 2 
with $I_{\rm rod}=7000$ A (i.e. $\beta=6.9$).}
\end{center}
\end{figure}

In the PROMISE 1 configuration, the upper endplate is a 
non-rotating Plexiglass lid P fixed to the frame F. 
The bottom, however, is simply part of the copper 
vessel V, and rotates with the outer cylinder. 
With respect to both their rotation rates and 
electrical conductivities, there is thus a clear asymmetry in the 
end caps. 

Figure 2a shows the changes made for PROMISE 2 
to improve this situation: First, both end caps 
are made of insulating material in order to avoid 
short-circuiting of currents along the copper end 
cap at the bottom (those currents had been shown to be 
dangerous by possibly changing the rotation profile via 
azimuthal forces \cite{JACEK1,JACEK3}). Second, both 
the upper and the lower end caps are split into two rings, 
the inner rotating with the inner cylinder and the outer 
rotating with the outer cylinder. In 
\cite{JACEK2} it had been shown that this splitting yields 
a minimization of the Ekman pumping if the position of the 
splitting is at 0.4 of the gap width $d$. Third, the co-rotation 
of the two rings with one of the cylinders made it necessary 
to change the signal path of the ultrasonic transducers. 

These two transducers provide full profiles of the  axial velocity 
along the beam-lines parallel to the axis of rotation. While in 
PROMISE 1 they are inserted in the upper Plexiglass end cap that 
is fixed to the laboratory frame, in PROMISE 2 they must be 
connected via a sliding contact to the signal processing computer 
(DOP 2000).

The configuration of magnetic fields is the same for PROMISE 1 and 
PROMISE 2. An axial magnetic field in the order of 10 mT is produced 
by a double-layer coil with 76 windings (C in Fig. 1). The 
omission of windings at two symmetric positions close to
mid-height, as seen in Fig. 1a, was motivated by a coil optimization 
to maximize the homogeneity of the axial field  throughout the 
fluid volume. The coil is fed by currents up to 200 A, beyond 
which a significant heating of the coil sets in. The azimuthal 
field, also in the  order of 10 mT, is generated by a current through 
a water-cooled copper rod R of radius 15 mm. The special power 
supply (PS in Fig. 1c) for this axial 
current is capable of delivering up to 8000 A.

\section{Some results}

We have carried out a large number of experimental runs 
in order to cover a wide range of parameter dependencies. 
Typically, the duration of an experimental run was 1900 
sec, after a waiting time of one hour and more. 
Many details for the PROMISE 1 
set-up can be found in the publications \cite{PRL,APJL,NJP,PROMISEAN}. 

\begin{figure}[t]
\begin{center}
\unitlength=\textwidth
\includegraphics[width=0.95\textwidth]{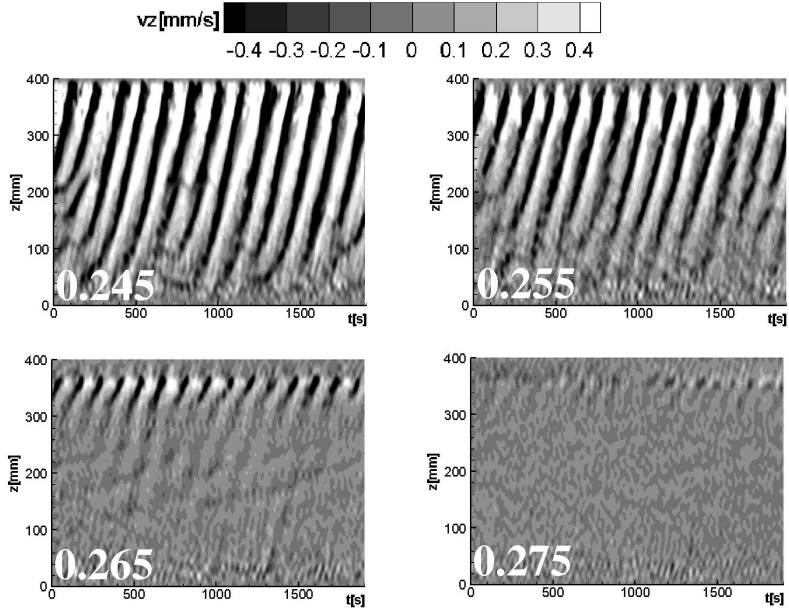}
\caption{Measured axial velocity perturbations for $\mu=$0.245, 0.255, 0.265, and 
0.275, at $f_{\rm i}=0.1$ Hz, $I_{\rm coil}=76$ A, $I_{\rm rod}=4000$A. }
\end{center}
\end{figure}

\begin{figure}[t]
\begin{center}
\unitlength=\textwidth
\includegraphics[width=0.95\textwidth]{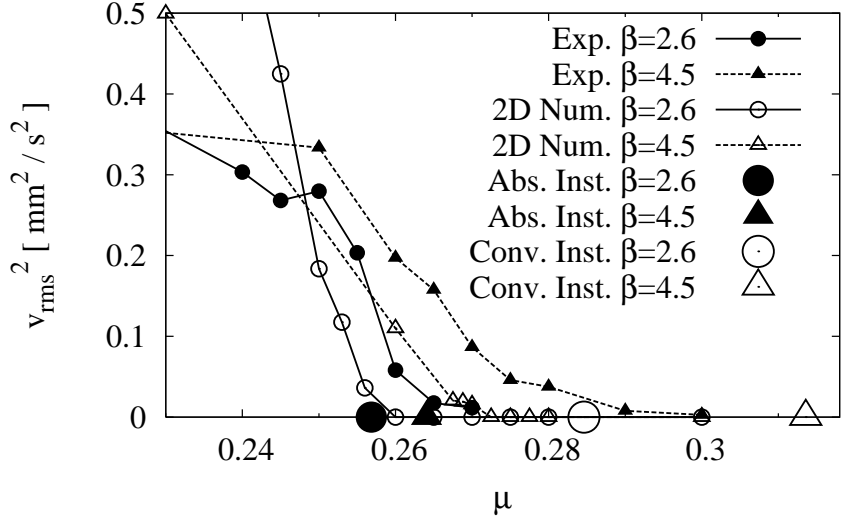}
\caption{Rms of the velocity perturbation for $f_{\rm i}=0.1$ 
Hz, $I_{\rm coil}=76$ A, and two values of $I_{\rm rod}=4000$ A and 7000 A, which 
corresponds to $\beta=2.6$ and $\beta=4.5$ respectively.
The experimental values (full lines) are compared with the numerical results of
a 2D solver, and with the results for the absolute and the convective instability
of a 1D solver (courtesy of J\=anis Priede).}
\end{center}
\end{figure}

One of the most significant features of the MRI is that, 
for fixed $Re$ and fixed azimuthal field $B_{\phi}$, it shows 
up only in a finite interval of the Hartmann number 
$Ha=B_{\rm z}(r_{\rm i} d \sigma/\rho \nu)^{1/2}$
($\sigma$ is the conductivity of the liquid, $\rho$ its density, and  
$\nu$ its kinematic viscosity). This appearance and 
disappearance of a travelling mode is a suitable indicator 
for the existence of the proper MRI mode and its distinction 
from other possible flow structures. 

In Fig. 2b,c we represent the results for
rotation rates of $f_{\rm i}=0.06$ Hz (i.e. $Re=1779$) 
and $f_{\rm o}=0.0162$ (i.e. $\mu=0.27$) which is 
slightly above  the Rayleigh value $\mu_{\rm Ray}:=r^2_{\rm i}/r^2_{\rm o}=0.25$.  
The current in the coil was fixed to
$I_{\rm coil}=50$A (i.e. Ha=7.9). In the PROMISE 1 
case (Fig. 2b), the axial current  
was set to $I_{\rm rod}=6000$ A (i.e. $\beta:=B_{\varphi}(r_{\rm i})/B_{\rm z}=5.9$), 
in the PROMISE 2 (Fig. 2c) case to $I_{\rm rod}=7000$ A (i.e. $\beta=6.9$), 
a difference that is, however, 
not essential. The grey scale of the plots 
indicates the axial velocity component $v_z$ measured along the 
ultrasound beam, from which we have subtracted the (z-dependent) 
time average in order to filter out the two Ekman vortices which 
appear already without any magnetic field. These Ekman vortices, 
which in the PROMISE 1 version are characterized by inward radial 
flows close to the upper and lower end-plates and a jet-like radial 
outflow in the centre of the cylinder \cite{KAGEYAMA}, are 
significantly suppressed in the PROMISE 2 version due to the 
use of split end caps. 
In PROMISE 1, the wave dies away at the position of the radial 
jet (which is not always at mid-height). 
In contrast to this, in PROMISE 2 the wave propagates throughout 
the total height of the cell.

This much improved behaviour of the MRI in PROMISE 2 can be
visualized nicely when crossing the Rayleigh line.
It is known \cite{NJP} that the travelling wave appears 
already with a stationary outer cylinder, i.e. at 
$\mu=0$, 
although with a very low frequency. With increasing $\mu$, 
the wave frequency 
increases and typically reaches a value of $(0.1....0.2)f_{\rm i}$
at the Rayleigh value 
$\mu_{\rm Ray}=0.25$. 
Figure 3 shows now in detail what happens with the travelling MRI 
wave
when $\mu$ crosses the Rayleigh line. 
The parameters for this run were
$f_{\rm i}=0.1$ Hz (i.e. $Re=2958$), $I_{\rm coil}=76$ A (i.e. 
$Ha=12$), $I_{\rm rod}=4000$ A. It is clearly visible that the MRI 
wave is still present at $\mu=0.255$, becomes significantly weaker at 
$\mu=0.265$, and has completely died at $\mu=0.275$.

This transition is further analyzed in Fig. 4 which 
shows the rms of the axial velocity perturbation
in dependence on $\mu$, now both for $I_{\rm rod}=4000$ A
and $I_{\rm rod}=7000$ A (which corresponds to 
$\beta=2.6$ and 4.5, respectively).

The measured values (full lines) are compared with the 
numerical results of
a 2D solver (dashed lines), but also with the results of a 1D solver 
for the onset 
of the
absolute and the convective instability \cite{JANIS1,JANIS2}
(courtesy of J\=anis Priede). It is most remarkable 
that the 
1D results for the onset of the 
absolute instability correspond nearly perfectly 
with the results of the 2D solver, while
the thresholds for the convective instability are situated 
at much higher values of $\mu$.
Compared to the 2D numerical curves, 
the experimental curves are
shifted only slightly towards higher $\mu$. As expected, we 
see also that for 
increasing $\beta$ the threshold of the 
instability shifts to higher values of $\mu$.

\section{Conclusions}

We have obtained experimental evidence for the existence of the 
MRI in current-free helical magnetic fields. 
The symmetrization of the axial boundary conditions 
and the use of split end caps in PROMISE 2 has led to a 
strong reduction of the Ekman pumping and hence to an avoidance 
of artefacts in the radial jet flow region. 
The MRI wave extends clearly beyond the Rayleigh line, 
and its behaviour is in good
correspondence with both 2D simulations and 1D simulations 
for the absolute instability, but in stark contrast with 
1D simulations for the convective instability.
This indicates that the observed MRI wave is indeed a global
instability and not only a noise triggered convective instability as
claimed recently \cite{LIU3}.
Further dependencies 
of the MRI on parameters like $Re$, $\beta$,  and $Ha$, as well 
as their comparison 
with numerical predictions will be published elsewhere.

\Thanks{We are grateful to J\=anis Priede for many fruitful discussions
on MRI, and for sharing with us his results for the 
convective and absolute instability. This work was supported by the
German Leibniz Gemeinschaft, within its ''Senatsausschuss Wettbewerb'' 
(SAW) programme.}



\begin{thebibliography}{xx}
\bibitem{VELIKHOV}
{\sc E.P. Velikhov}.
\newblock {Stability of an ideally conducting liquid 
fluid between cylinders rotating in a magnetic field}.
\newblock {\it {Sov. Phys. JETP}\/}, vol.~9 (1959), pp.~995--998.
\bibitem{BAHA}
{\sc S.A. Balbus and J.F.   Hawley}.
\newblock {A powerful local shear instability in weakly magnetized disks. 
1. Linear analysis}.
\newblock {\it {Astrophys. J.}\/}, vol.~376 (1991), pp.~214--222.
\bibitem{RMP}
{\sc A. Gailitis, O. Lielausis, E. Platacis, G. Gerbeth, and F. Stefani}.
\newblock {Colloquium: Laboratory experiments on hydromagnetic dynamos}.
\newblock {\it {Rev. Mod. Phys.}\/}, vol.~74 (2002), pp.~973--990.
\bibitem{ZAMM}
{\sc  F. Stefani, A. Gailitis, and G. Gerbeth}.
\newblock {Magnetohydrodynamic experiments on cosmic magnetic fields}.
\newblock {\it {ZAMM}\/}, 88 (2008), 930--954.
\bibitem{SISAN}
{\sc D. Sisan et al.}.
\newblock {Experimental observation and characterization of the magnetorotational instability}.
\newblock {\it {Phys. Rev. Lett.}\/}, vol.~93 (2004), Art. No. 114502.
\bibitem{HORU}
{\sc R. Hollerbach and G. R\"udiger}.
\newblock {New type of magnetorotational instability in cylindrical Taylor-Couette flow}.
\newblock {\it {Phys. Rev. Lett.}\/}, vol.~95 (2005), Art. No. 124501.
\bibitem{PRL}
{\sc F. Stefani et al.}.
\newblock {Experimental evidence for magnetorotational instability 
in a Taylor-Couette flow under the influence of a helical magnetic field}.
\newblock {\it {Phys. Rev. Lett.}\/}, vol.~97 (2006), Art. No. 184502 .
\bibitem{APJL} 
{\sc G. R\"udiger et al.}.
\newblock {The travelling-wave MRI in cylindrical Taylor-Couette flow: 
Comparing wavelengths and speeds in theory and experiment}.
\newblock {\it {Astrophys. J.}\/}, vol.~649 (2006), pp. L145-L147.
\bibitem{NJP}
{\sc F. Stefani et al.}.
\newblock {Experiments on the magnetorotational instability in helical  magnetic fields}.
\newblock {\it {New J. Phys.}\/}, vol.~9 (2007), Art. No. 295.
\bibitem{PROMISEAN}
{\sc  F. Stefani et al.}.
\newblock {Results of a modified PROMISE experiment}.
\newblock {\it {Astron. Nachr.}\/}, vol.~329 (2008), pp.~652-658.
\bibitem{JACEK1}
{\sc   J. Szklarski}.
\newblock {Ekman-Hartmann layer in a magnetohydrodynamic Taylor-Couette flow}.
\newblock {\it {Phys. Rev. E}\/}, vol.~76 (2007), Art. No. 066308.
\bibitem{JACEK3}
{\sc  J. Szklarski and G. Gerbeth}.
\newblock {Boundary layer in the MRI experiment PROMISE}.
\newblock {\it {Astron. Nachr.}\/}, vol.~329 (2008), pp.~667-674.
\bibitem{JACEK2}
{\sc  J. Szklarski}.
\newblock {Reduction of boundary effects in spiral MRI experiment PROMISE}.
\newblock {\it {Astron. Nachr.}\/}, vol.~328 (2007), pp.~499--506.
\bibitem{KAGEYAMA}
{\sc  A. Kageyama, H. Ji, J. Goodman, F. Chen, E. Shoshan}.
\newblock {Numerical and experimental investigation of circulation in short cylinders}.
\newblock {\it {J. Phys. Soc. Jpn.}\/}, vol.~73 (2004), pp.~2424-2437.
\newblock {\it {Astron. Nachr.}\/}, vol.~329 (2008), pp.~659-666.
\bibitem{JANIS1}
{\sc  J. Priede, I. Grants, and G. Gerbeth}.
\newblock {Inductionless magnetorotational instability in a Taylor-Couette flow with a helical magnetic field}.
\newblock {\it {Phys. Rev. E}\/}, vol.~75 (2007), Art. No. 047303.
\bibitem{JANIS2}
{\sc  J. Priede and G. Gerbeth}.
\newblock {Absolute versus convective helical 
magnetorotational instability in a Taylor-Couette flow}.
\newblock {\it {Phys. Rev. E}\/}, submitted (2008); arxiv:0810.0386
\bibitem{LIU3}
{\sc  W. Liu}.
\newblock {Noise-sustained convective 
instability in a magnetized Taylor-Couette flow}.
\newblock {\it {Astrophys. J}\/}, submitted (2008); arxiv:0808.2513
\end{thebibliography}
\end{document}